\documentclass[preprint,12pt]{elsarticle}
\usepackage[dvips]{epsfig}
\usepackage{psfrag}
\usepackage{amssymb,amsmath}
\usepackage[dvips]{graphicx}
\usepackage{subfig}
\usepackage[usenames,dvipsnames]{color}
\usepackage{float} 

\begin{document}

\begin{frontmatter}

\title{Modeling Rectification Effects in Mesoscopic Superconducting Devices}

\author[MFC]{{M. F. Carusela}}
\cortext[MFC]{Corresponding author}
\ead{flor@ungs.edu.ar}

\author[VR]{V. P. Ramunni}
\ead{vpram@cnea.gov.ar}

\author[VM]{V. I. Marconi}
\ead{vmarconi@famaf.unc.edu.ar}


\address[MFC]{Instituto de Ciencias, UNGS, J.M. Guti\'errez 1150,\\ 1613 Los Polvorines, Argentina - CONICET.\\}
\address[VR]{Departamento de Materiales, CAC-CNEA, Avda. General Paz 1499,\\ 1650 San Mart{\'i}n, Argentina - CONICET.}
\address[VM]{Facultad de Matem\'atica, Astronom{\'i}a y F{\'i}sica,   Universidad Nacional de C\'ordoba and  IFEG-CONICET, X5000HUA C\'ordoba, Argentina.\\}

\begin{abstract}

We study thermal fluctuations and capacitive effects  on small Josephson Junction Rings (JJR) that mimics  
the rectification phenomena recently observed in triangle shaped mesoscopic superconductors,  due to the superposition of the field induced persistent current with the bias current. 
At finite temperature we predicted that the amplitude of the rectified signal depends strongly on the current contacts configuration
on the JJR, in coincidence  with experiments.
In addition we analize the range of parameters where a closed loop of capacitive junctions is an appropriate model to explain the experimental observations.
We conclude that the closed loop of weak links, a JJR,   is a simple, robust and good enough model to explain the observed voltage rectification effects on mesoscopic superconducting samples for a wide range of temperature.

\end{abstract}


\end{frontmatter}

\section{INTRODUCTION}

Technological  progress  
allowing the manipulation of ratchet effects in complex systems 
such as cold atoms \cite{gommers06},  colloids \cite{colloids},  cells \cite{biology}, fluids \cite{matthias},  electrons in semiconductors \cite{linke}, extended magnetic walls \cite{perez},  vortices in Josephson systems \cite{falo99,shalom}, motivate this work.
In  all the mentioned examples the flux of driven particles interact with an asymmetric potential. 
The key question on this field is that {\it it is not simple} in each case to know feasibly which are the mechanisms underlying these many bodies rectification effects.
It is well known that the velocity of a single overdamped particle under
  an applied ac force can be rectified by introducing a periodic potential
  with broken reflection symmetry \cite{rat_rev}. We can thus ask 
  whether this effect, known as rocking  ratchet (because the system is rocked with an external ac force), 
  can be still observed as a rectified dc voltage response in transport measurements  
 in singly connected mesoscopic superconductors under external ac bias currents, thermal fluctuations and external homogenous magnetic field applied. Indeed finite dc voltage was measured under zero averaged ac current applied and it was extensively studied in superconducting (SC) nanostructures \cite{nele0, nele, neleth}.
But  this system differs from examples mentioned previously, because the essential broken symmetry required to obtain a rectifier is controlled differently in these samples, it is controlled by its geometry (sample shape) and the configuration of contacts leads (position of the current/voltage probes).  
In particular, motivated by these recent experiments reporting original voltage rectification effects in mesoscopic superconducting triangles \cite{nele} we simulated a more realistic  model for such  system in order to understand and corroborated  the physics behind the observed ratchet effects, with special     emphasis in the finite temperature effects not taking into account in previous and simpler models. 

 In a pionner analytical work \cite{brandt}  different asymmetric type-II superconducting structures (rings and strips) were already proposed and analyzed under current and magnetic field applied. They show how  magnetic flux penetration depending on the geometric parameters is fieldlike or currentlike. Later on,  measurements were realized in asymmetric SC rings \cite{dubonos, dubo-cm}. In addition similar rectification effects were reported in a singly connected structured if the current injection is off-center \cite{nele0}. From mentioned works it is  widely accepted that the effects observed are due to the asymmetry causing a difference in critical current for a positive or negative applied external current which is compensated or reinforced by the field induced persistent current. Further an important parameter, temperature, should be taking into account to model  in detail the phenomena observed. First because  all results reported in single triangle  samples are done close to $T_c$ \cite{nele0, nele, neleth}, it means where thermal fluctuations are important and far away from $T=0$ where previous theoretical analysis was done \cite{nele}. In addition it is well known that geometry in microsized or in nanosized samples could play essential roles in fluctuations phenomena and even more, very recently it was shown that the superconducting order parameter present larger thermal  fluctuations in the corners of sharp  shaped samples \cite{pogosov}. At present,  when smaller dimensions are manipulated it is motivating to study effects of thermal fluctuations even at temperature much lower than $T_c$  where dimensionality start to compete and the limits between thermal fluctuations and quantum fluctuations are not clear. In short, to understand how thermal fluctuations influence the rectification phenomena at all  finite temperature is important. 

In this paper we model voltage rectification effects observed in thin mesoscopic superconducting triangles \cite{nele} using two kind of  small Josephson Junction Rings: a resistive one,  using  the RSJ (resistively  shunted junction) model  and a capacitive model, using the RCSJ (resistevely and capacitive shunted junction) model. We introduce finite temperature effects adding white noise to the resulting dynamics equations within the RSJ model. We find   good qualitative agreement with experiments.

\section{SIMULATED MODEL}
\begin{figure}[H]
\begin{center}
\includegraphics[width=12cm]{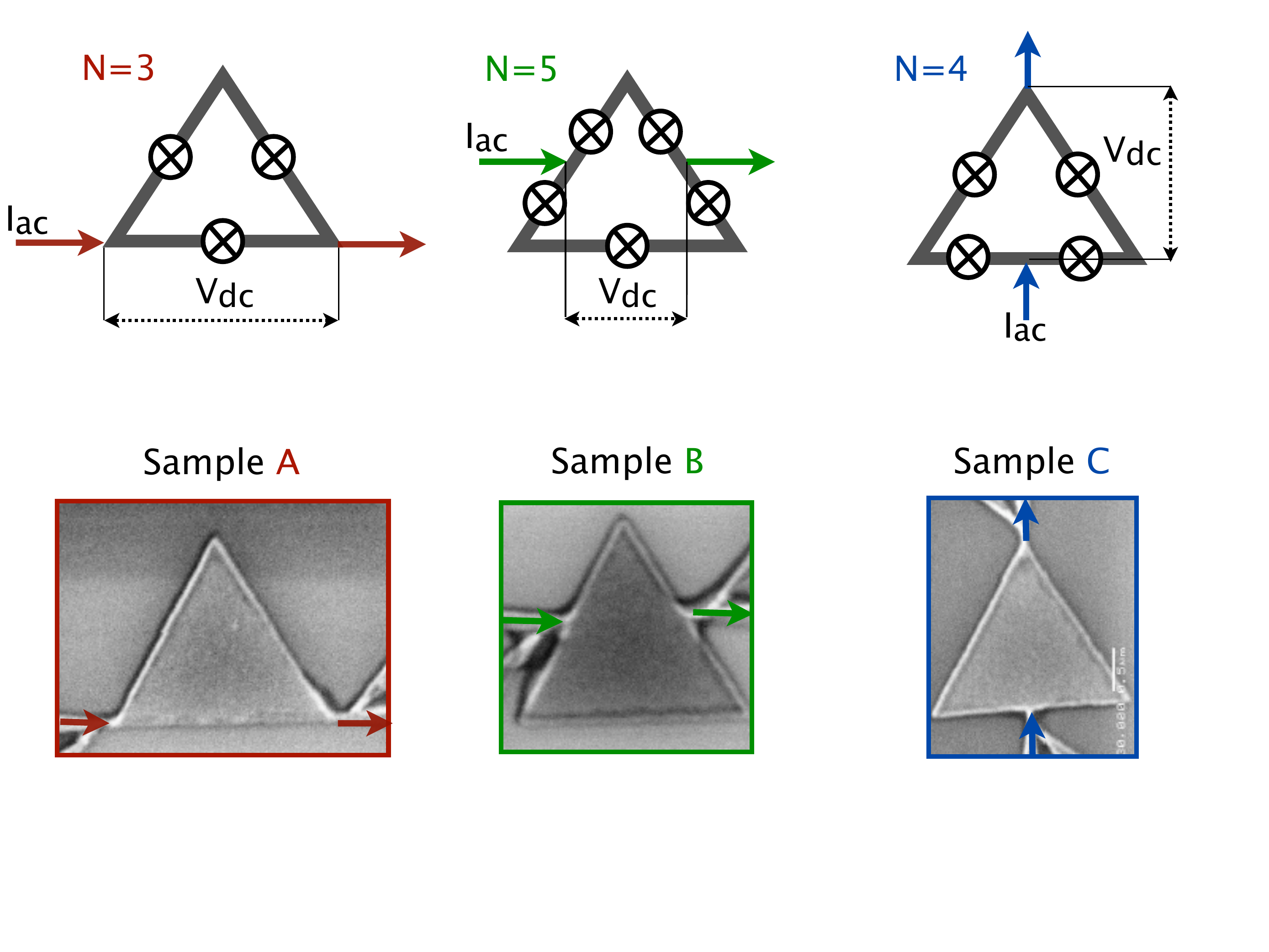}
\end{center}
\caption{(Color online) Schematic small Josephson junctions rings modeling superconducting triangles, using either $N=3$ JJ, $N=5$ JJ or $N=4$ JJ (upper panels), according  to the  experimental samples A, B or C with different contact positions (lower panels). Note the different asymmetric current injection for $N=3$, A,  below the geometrical center and $N=5$, B, above the geometrical center  in opposite to the symmetric current injection for $N=4$, C. Lower panels (generosity of Nele Schildermans and A.V. Silhanek) are scannig electron microscopy images of the superconducting Al triangles used in experiments.} 
\label{fig:jjr}
\end{figure}

 We study at finite temperature the dynamics of  small Josephson Junction Rings (closed loops) with  a finite number of identical Josephson junctions (JJ): three JJ $N=3$, four JJ $N=4$  or five JJ, $N=5$ as sketched in  Fig.\ref{fig:jjr} for different positions of the current injection.  
 An external sinusoidal ac current  is applied  in different loop points  as shown   in Fig.\ref{fig:jjr} in order to mimics the different contact configurations, samples A, B and C used in experiments \cite{nele}.

\begin{figure}[H]
\begin{center}
\includegraphics[width=12cm]{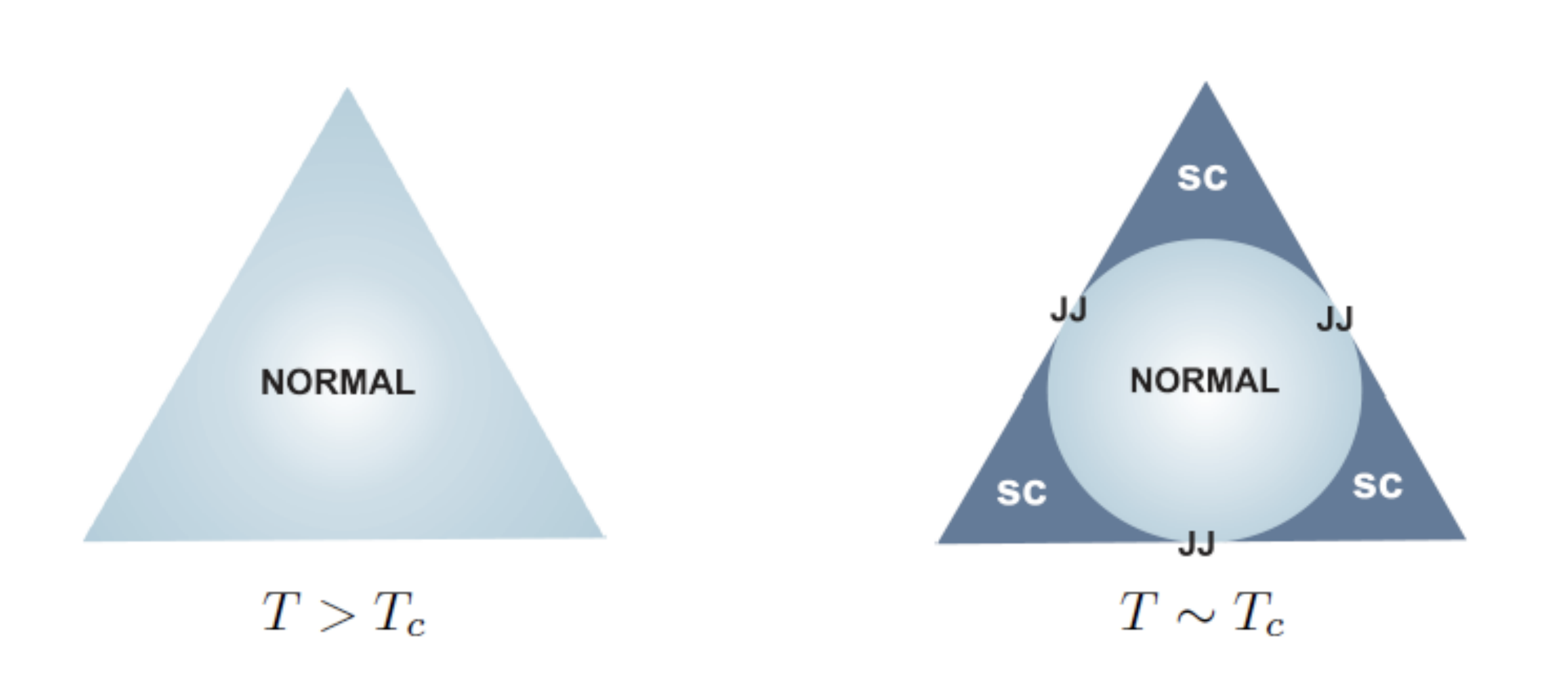}
\end{center}
\caption{(Color online) At large tempererature  the Al triangles are normal samples but at $ T \sim T_c$ nucleation of the superconducting condensate starts from the sharp sample corners.  The order parameter $\Psi$ is maximum at the triangle vertices and minimum at the middle of the sides, then a ring  of weak links (JJ) is used to model this system.} 
\label{fig:tri}
\end{figure}

 In Fig.\ref{fig:tri}  we show a schema of the shape in which superconductivity nucleation occurs  in triangle shaped mesoscopic samples.  It is well known that nucleation in mesoscopic samples is dominated by the surface superconductivity and its shape and boundaries, {\it i.e} by its geometry. In fact, it was clearly shown that surface nucleation is enhanced in wedge shaped superconductors \cite{fomin, peeters}, then the idea to model the triangle as a loop of  strong superconducting electrodes (SC islands) at the triangle corners  connected by weak links is reasonable. Those weak links, could be an insulator  as Josephson proposed originally (SIS JJ) or a normal metal (SNS JJ), or simpler a short narrow constriction of the same superconductor (ScS JJ) \cite{tink}.  
Our main assumption for the calculations is that the
superconducting order parameter $\rho({\bf r})=|\rho({\bf
r})|e^{i\theta({\bf r})}$ is such that $|\rho({\bf r})|=\rho_0$
with $\rho_0$ the same constant on all SC islands and $\theta({\bf r})$
is spatially constant in each island. Then our variables will be the superconducting phases in each SC island.

 We started solving the simpler case,  the ring of weak links of SNS junctions, using  the resistively shunted junction (RSJ) model plus thermal fluctuations and solving numerically the superconducting phase
dynamics using Langevin dynamics technique \cite{dyna}. We  considere a total magnetic field $B$ applied perpendicular to the ring,  spatially and
temporally constant. The  Hamiltonian of a closed loop of $N$ SNS-junctions is the following:
\begin{equation}
 H = -E_J \sum_{n=0}^{N-1} \cos(\phi_n - a_n)
\end{equation}
where $E_J$ is the Josephson energy, $\phi_n = \theta({\bf r}_n) - \theta({\bf r}_{n-1})$ is the superconducting phase
difference at the junction $n=0,...,N$, and $\theta({\bf r}_n)$ is the phase of
the superconducting island centered at 
\begin{equation}
{\bf r}_n =  R\left(-\cos\left(\frac{2\pi n}{N} \right) {\hat x} + \sin\left(\frac{2\pi n}{N}\right) {\hat y}\right)
\end{equation}
being $R$ the ring
radius. Note here that our loops imitating the triangle samples, are ideal junction circles where the needed triangle asymmetry can be introduce controlling the number of junctions and the points where external currents are injected and extracted (see Fig.~\ref{fig:jjr}).  
The total magnetic field is ${\bf B}=\nabla
\times {\bf A}$.
The magnetic field contribution to the phase difference $a_{n}$ is
the line integral of the vector potential between sites $n$ and
$n-1$,
\begin{equation}
 a_n = \frac{2\pi}{\Phi_0} \int_{{\bf r}_{n-1}}^{{\bf r}_{n}}{\bf A}({\bf l}).d{\bf l}, 
\end{equation}
being $\Phi_0=h/2e$ the single superconducting quantum flux. 
After taking the magnetic field applied perpendicular to the sample plane, ${\bf B}=B {\bf z}$, the invariant gauge ${\bf A}= B x {\hat
y}$ for the vector potential ${\bf A}$  and  the flux quantum number
through the ring $\Phi/\Phi_0 = \pi R^2 B/\Phi_0$ we obtain after solving the corresponding previous integral in polar coordinates:
\begin{equation}
 a_n =   -\frac{\Phi}{2\Phi_0} \biggl[\frac{4\pi}{N} + \sin\left(\frac{4\pi n}{N}\right)- \sin\left(\frac{4\pi(n-1)}{N}\right)\biggl].
\label{eq:an}
\end{equation}

The current flowing in the JJ between two superconducting islands is modeled by a typical electric circuit build up of an ideal JJ    in  parallel with a normal resistance, {\it i.e} by the  sum of the Josephson supercurrent and the normal electron current. 
Conservation of current is considered with Kirchhoff laws  fulfilled in each node. We inject a current $I$ between
junctions $N-1$ and $0$, and extract it $\delta$ junctions away,
between junctions $\delta-1$ and $\delta$. The resulting set of
dimensionless equations for the currents flowing in the ring is
the following:

\begin{equation}
\dot{\phi}_n =  I_{up} - \sin(\phi_n - a_n)  + \Gamma(n,t), \;\;\;\;\;     0 \le n \le
\delta-1\\
\label{eq:lan1}
\end{equation}
 
 \begin{equation}
\dot{\phi}_n = I_{up}-I -  \sin(\phi_n - a_n) + \Gamma(n,t), \;\;\;\;\;  \delta \le n
\le N-1 \\
\label{eq:lan2}
\end{equation}

\begin{equation}
I_{up} \equiv (1-\delta/N) I + \frac{1}{N} \sum_{n=0}^{N-1} \sin(\phi_n
- a_n) + \Gamma(n,t).
\label{eq:iup}
\end{equation}

which are $N$ first order differential equations for the time
evolution of the $N$ phase variables $\{\phi_n\}_{n=0}^{N-1}$. $I_{up}$ is the current flowing in the upper branch of the ring, up with respect to the injection and extraction current points, and $I-I_{up}=I_{down}$. Let
us note that each junction interacts with all the others through
$I_{up}(\{\phi_n\}_{n=0}^{N-1})$, the total current in the upper
branch of the circuit, which represents a kind of mean-field
interaction plus a drive. 
In addition finite temperature effects are taking into account through an added   Langevin
noise term $\Gamma$,  introduced as current fluctuations,   which  models the contact with a thermal bath at temperature
$T$ and   satisfies the condition for non-correlated white noise:

\begin{equation}
  \langle \Gamma( n,t)\Gamma(n^{\prime },t^{\prime })\rangle=\frac{2k_BT}{R_{N}}\delta _{n,n^{\prime }}\delta (t-t^{\prime }). 
\end{equation}

where $R_N$ is the normal state resistance.  In short the model control parameters are the external current $I$, magnetic field applied or magnetic flux $\Phi/\Phi_0$ and temperature T.  In addition
one can play with the source-drain distance $\delta$ and the number of junctions
N.

 Langevin dynamical equations Eq.~\ref{eq:lan1}-\ref{eq:lan2} using conditions (\ref{eq:an}) and (\ref{eq:iup}) are solved
numerically using a second order Runge-Kutta-Helfand-Greenside algorithm under an external sinusoidal current applied $I=I_{ac}\sin(2\pi\omega t)$. We calculate the   mean voltage response computing the instantaneous dc voltage drop $v$ between source and drain:

\begin{equation}
v =  \sum_{n=0}^{\delta-1} \dot{\phi}_n =  \delta I_{up} - \sum_{n=0}^{\delta-1} \sin(\phi_n - a_n) +  \Gamma(n,t).
\end{equation}

In fact there is a second equivalent form to compute the voltage drop, if  upper and lower  paths contain a different number of junctions:

\begin{equation}
 v = - \sum_{n=\delta}^{N-1} \dot{\phi}_n =  (N-\delta) (I-I_{up}) + \sum_{n=\delta}^{N-1} \sin(\phi_n - a_n) +  \Gamma(n,t).
\end{equation}

We normalize currents by the single junction
critical current $I_0$,      
voltages by $R_N I_0$ , temperature by $E_J/k_B$ and the unit of time is $\tau_J$=$2\pi cR_N I_0 /\Phi_0$. 

Possibly weak links sketched in Fig.~\ref{fig:tri} could be more complex than the simpler SNS junctions, like tunnel  superconducting-insulator-superconducting (SIS) junctions and a more complete description is required. Then we probe if the essential features of previous experiments \cite{nele0,nele} are as well described by a RCSJ  (resistively and capacitively shunted junction, \cite{tink}) ring. In this model the weak link (JJ) is modeled by an ideal JJ shunted in parallel by a resistance $R_c$ and a capacitance $C$. The addition of capacitive effects accounts for the geometric shunting capacitance between the two superconducting islands. Now the following set of second order differential equations should be solved:

\begin{equation}
 \ddot{\phi}_n + \eta \dot{\phi}_n + \sin(\phi_n - a_n) =  I_{up}, \;\;\;\;\;        0 \le n \le
\delta-1 
\end{equation}
 
 \begin{equation}
 \ddot{\phi}_n +  \eta \dot{\phi}_n + \sin(\phi_n - a_n) = I_{up}-I, \;\;\;\;\;  \delta \le n
\le N-1 
\end{equation}

being $\eta$ the main parameter now, a measurement of the system dissipation, $\eta=1/\sqrt{\beta}$ and $\beta$ is known as the McCumber parameter \cite{beta}, or damping parameter equal to $(\omega_pR_cC)^2$ with $\omega_p$ the plasma frequency of the junction $\omega_p = \sqrt{2e I_0/ \hbar C }$. 
\begin{figure}[ht]
\begin{center}
\includegraphics[height=9.5cm]{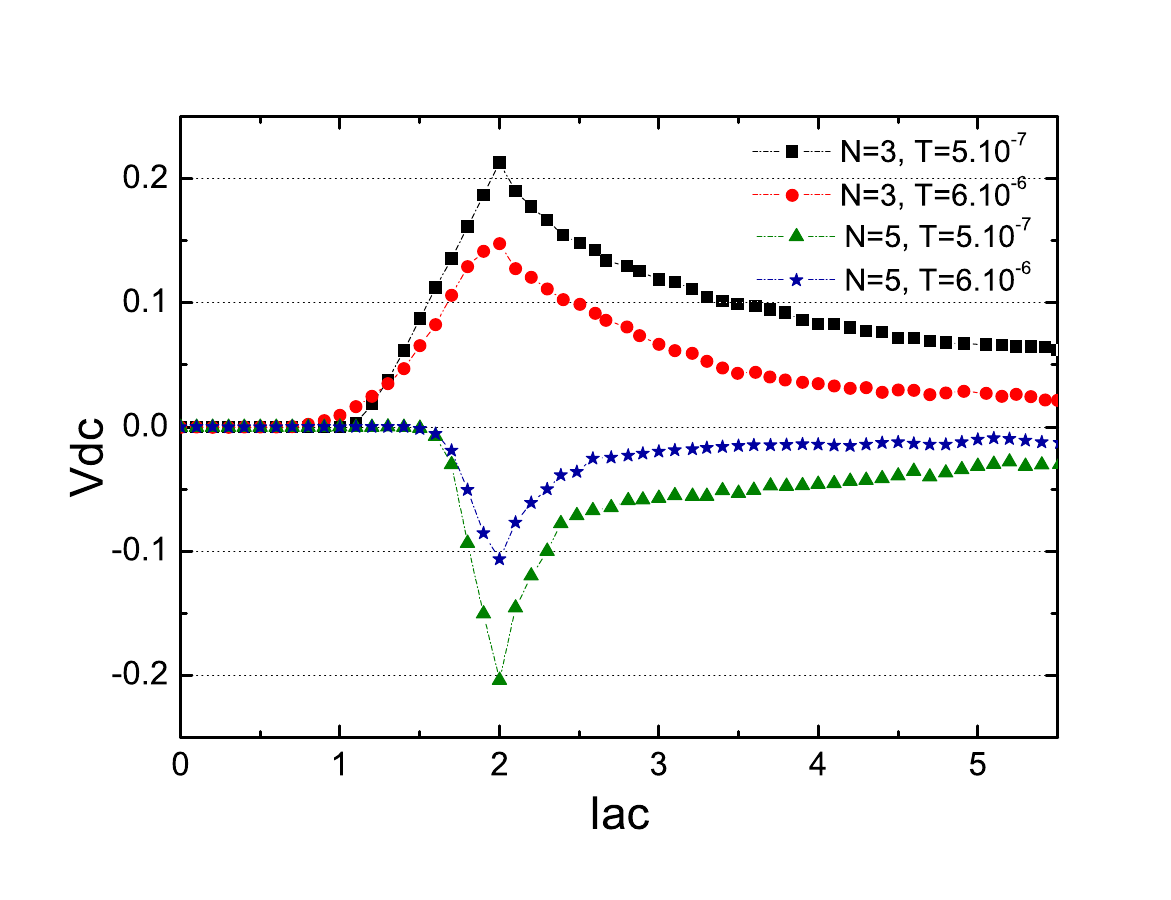}
\end{center}
\caption{ (Color online) Rectified voltage vs ac-current amplitude  applied for $N=3$ (sample A) and $T=5 \times 10^{-7}$(black $\color{black} \blacksquare\color{black})$, $N=3$ and $T=6 \times 10^{-6}$(red  $\color{red}\bullet\color{black}$),  $N=5$ (sample B) and  $T=5 \times 10^{-7}$(green  $\color{green}\blacktriangle\color{black}$), $N=5$ and $T=6 \times 10^{-6}$ (blue  $\color{blue}\bigstar\color{black}$). For all examples shown: $\Phi/\Phi_{0} = -1/4 $. The symmetric sample C, $N=4$, it is not shown for clarity because $V_{dc}=0$ for all $I_{ac}$  and all magnetic field applied.}
\label{fig:ivs}
\end{figure}
\section{RESULTS}

In order to study the influence of contacts on the rectification effects  observed in superconducting triangles \cite{nele}   the model for RSJ  rings described in previous section was first solved numerically. We calculated the rectified mean dc voltage $V_{dc} = \langle v
\rangle$ as a function  of the
ac current amplitude,  $I_{ac}$,  in the
low frequency limit  at finite temperatures and for different magnetic fields applied perpendicularly to the samples. 
The results for closed loops  with $N=3$  junctions (which mimics the experimental contacts configuration of sample A) and $N=5$ (sample B), both with
the same source-drain distance $\delta=2$, are shown in
Fig.~\ref{fig:ivs}.

We can clearly see that at {\it finite temperature} both, the $N=3$ and $N=5$ devices  display  finite rectified voltage,  $i.e.$ $|V_{dc}|>0$ above a critical current. Examples for a fixed magnetic field corresponding to  $n=-1$, {\it i.e.} for $\Phi/\Phi_{0} = -1/4$ and two temperature values are shown. The junction loop with $N=5$ presents a clear difference comparing with the case $N=3$, {\it the rectified voltage is inverted}, $i.e.$ $V_{dc} < 0$, resulting both junction loops   in good choices to reproduce the experimental results obtained in Ref.~\cite{nele} for both  type of asymmetric contact configurations, sample A and sample B, for which inverted rectified voltage responses were reported. In short, this simple model including thermal fluctuations mimics correctly recent experimental results \cite{nele}. 
Other details could be extracted from Fig.~\ref{fig:ivs}, 
all examples shown present a critical current amplitude 
threshold that decreases with temperature and it  is larger for the device with the larger number of junctions, $N=5$.  
We can also conclude that the
maximum of $|V_{dc}|$ is almost the same in both cases when the temperature is closer to zero, but when temperature is increased these maximum decrease, as it is expected when thermal fluctuations are present, but the  rates are different depending on the sample type, the maximum rectified voltage $|V_{dc}|_{max}$ decreases quicker in the larger loop, $N=5$. At $T= 5  \times 10^{-7}$ both cases present a maximum at  $|V_{dc}|_{max} \sim 0.21$ but when temperature is decreased more than an order of magnitude, 
$T=6 \times 10^{-6}$, the maximum is  $|V_{dc}|_{max} \sim 0.15$ for $N=3$ and   $|V_{dc}|_{max} \sim 0.1$ for $N=5$. In addition $|V_{dc}|$ 
  decays to zero  with $I_{ac}$, as expected, but this behavior is quicker for $N=5$.

\begin{figure}[ht]
\begin{center}
\includegraphics[height=9.5cm]{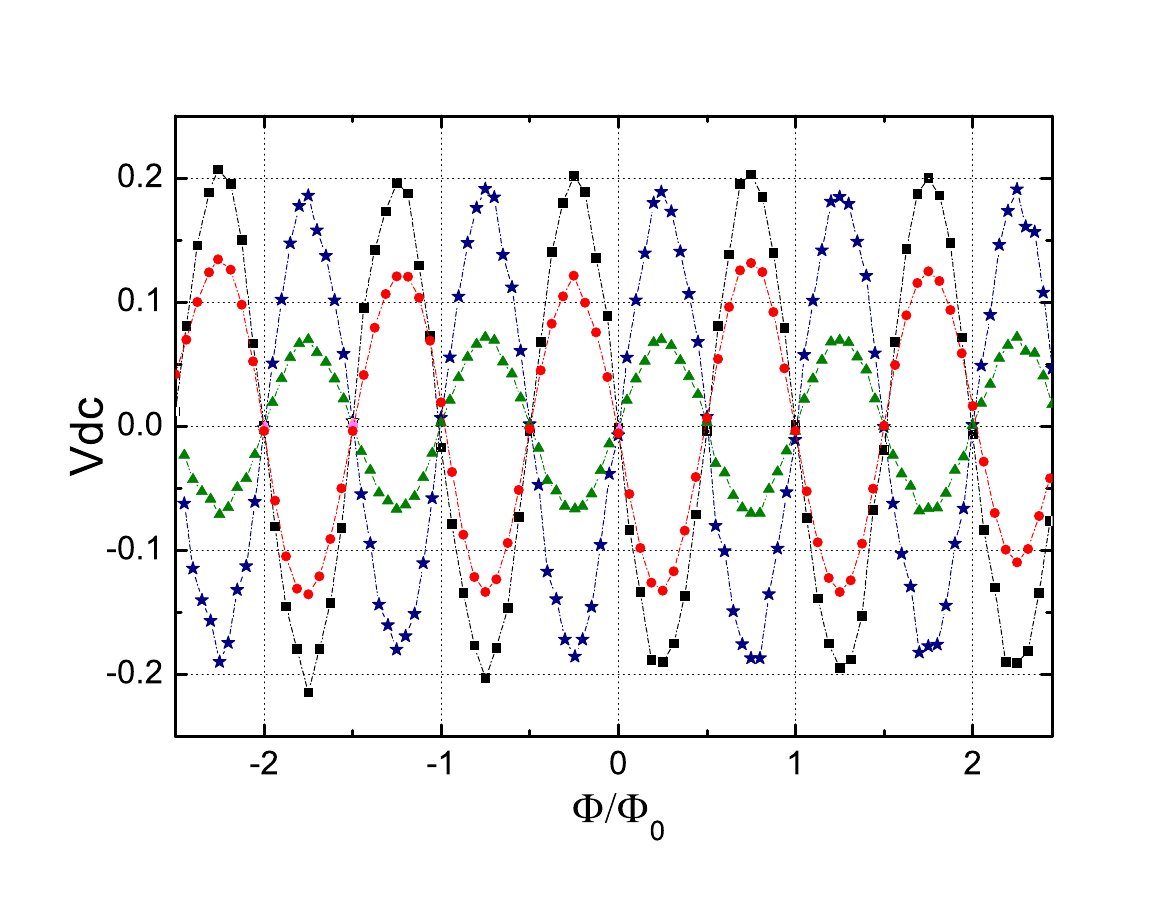}
\end{center}
\caption{(Color online) Oscillations of the rectified dc-voltage  $V_{dc} $  vs  magnetic flux $\Phi/\Phi_{0}$, comparing both samples and two finite temperatures:   $N=3$ and  $T=5 \times 10^{-7}$(black $\color{black} \blacksquare\color{black})$, $N=3$ and $T=1 \times 10^{-5}$(red  $\color{red}\bullet\color{black}$),  $N=5$  and  $T=5 \times 10^{-7}$  (blue  $\color{blue}\bigstar\color{black}$), $N=5$ and $T=1 \times 10^{-5}$ (green  $\color{green}\blacktriangle\color{black}$).}
\label{fig:vphi}
\end{figure}
\begin{figure}[ht]
\begin{center}
\includegraphics[height=9.5cm]{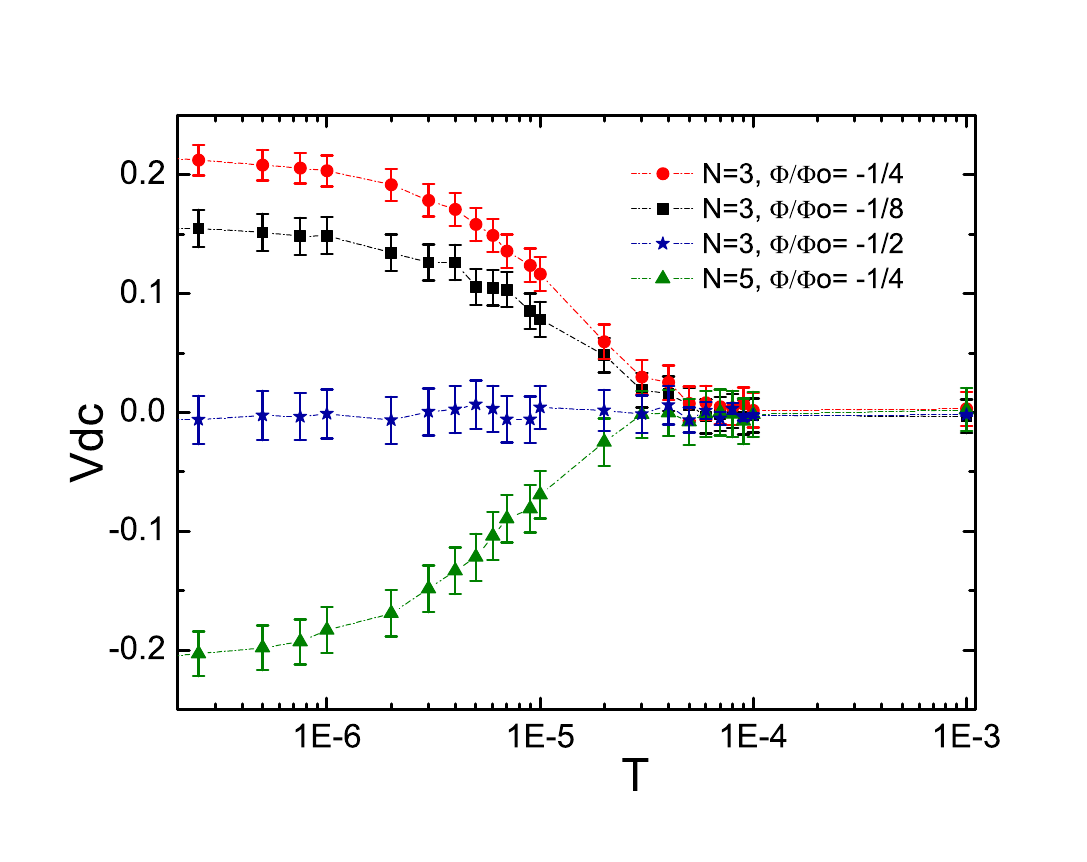}
\end{center}
\caption{(Color online) Rectified dc voltage $V_{dc}$ vs temperature $T$  in log-normal scale for $I_{ac}=2$ in all examples and  $N=3$ and $\Phi/\Phi_{0}=-1/4$ (red $\color{red} \bullet\color{red}$), $N=3$ and $\Phi/\Phi_{0}=-1/8$ (black  $\color{black}\blacksquare\color{black}$), $N=3$ and $\Phi/\Phi_{0}=-1/2$ (blue $\color{blue}\bigstar\color{blue}$), $N=5$ and $\Phi/\Phi_{0}=-1/4$ (green $\color{green}\blacktriangle\color{black}$).}
\label{vtemp}
\end{figure}

Another interesting feature to analize is the voltage response related to the total magnetic field applied, $B$.
 An important behavior  is obtained and analyzed in Fig.\ref{fig:vphi}, where  qualitatively the same
oscillations  for $V_{dc}$ as a function of vorticity (magnetic flux  $\Phi/\Phi_{0}$ through the ring) are also
obtained in the experiments \cite{nele}. Three characteristics of the oscillations dependency on magnetic field  are found to be independent of temperature: (a)  the rectification effects disappear if $ \Phi/\Phi_0
= n/2$, being $n$  an integer,  (b) the maximum absolute value of the rectified voltage appears always the condition $\Phi=(2n+1)\Phi_0/4$ is fulfilled and  (c) $|V_{dc}| \neq 0$ if $ \Phi/\Phi_0
\neq n/2$.
The observed dependences of rectified voltage on magnetic field are in accordance with expected Little-Parks oscillations, which occur every time the vorticity of the system increase due to the entrance of a new vortex: $L \rightarrow L+1$  \cite{little}.  In addition a noticeable feature appears, the sign of the rectified voltage changes, from  negative if $n < \Phi/\Phi_0 < n + 1/2$  to positive if $n + 1/2 < \Phi/\Phi_0 < n + 1 $ (being $n$ an integer) for $N=3$ and signs are inverted   for  $N=5$. This opposite behavior between both devices is related with its asymmetry and it is the best evidence that indeed the observed rectification effects are due to the superposition of the external applied sinusoidal current (bias current) with the field induced persistent current circulating around the JJRs. The external current injection is well below the geometrical center for  $N=3$ device on the contrary to $N=5$ device, then an opposite current compensation behavior was obtained. It is worth to mention that all simulations were performed as well for the symmetric current injection case, sample C, and no voltage response was observed, corroborating the essential need of an off-center current injection (system asymmetry induced externally for the contacts configuration).
 Note that $|V_{dc}|$ decreases to zero while temperature is increased,   in both cases.
 In order to analyze this last feature in detail comparing both contact configurations, the rectified dc voltage  as a function of  temperature is shown  in Fig.~\ref{vtemp} for  $I_{ac}=2$  and some typical values of the magnetic field applied, corresponding to $\Phi/\Phi_{0} = (2n+1)/4=-1/4$ (maximum voltage response), $\Phi/\Phi_{0}= (4n+1)/8 = -1/8$ (intermediate voltage response) and $\Phi/\Phi_{0}=-1/2$ (minimum voltage response).  For fields with finite response, the absolute value of the rectified voltage decrease smoothly to zero with similar temperature functionality  while temperature is incremented slowly in  several orders of magnitude.  It is worth to mention that
  it remains finite for a wide range of temperature. Analyzing the three curves presented for $N=3$,  a  rectified voltage tendency   to zero  as the magnetic flux approach to $n/2$ is observed as well, in coincidence with previous Fig.~\ref{fig:vphi}. These results show clearly the robustness of the rectification effects under a wide range of thermal fluctuations 
 for all magnetic field applied such that $\Phi/\Phi_{0} \neq n/2$. 
 \begin{figure}[ht]
\begin{center}
\includegraphics[height=9.cm]{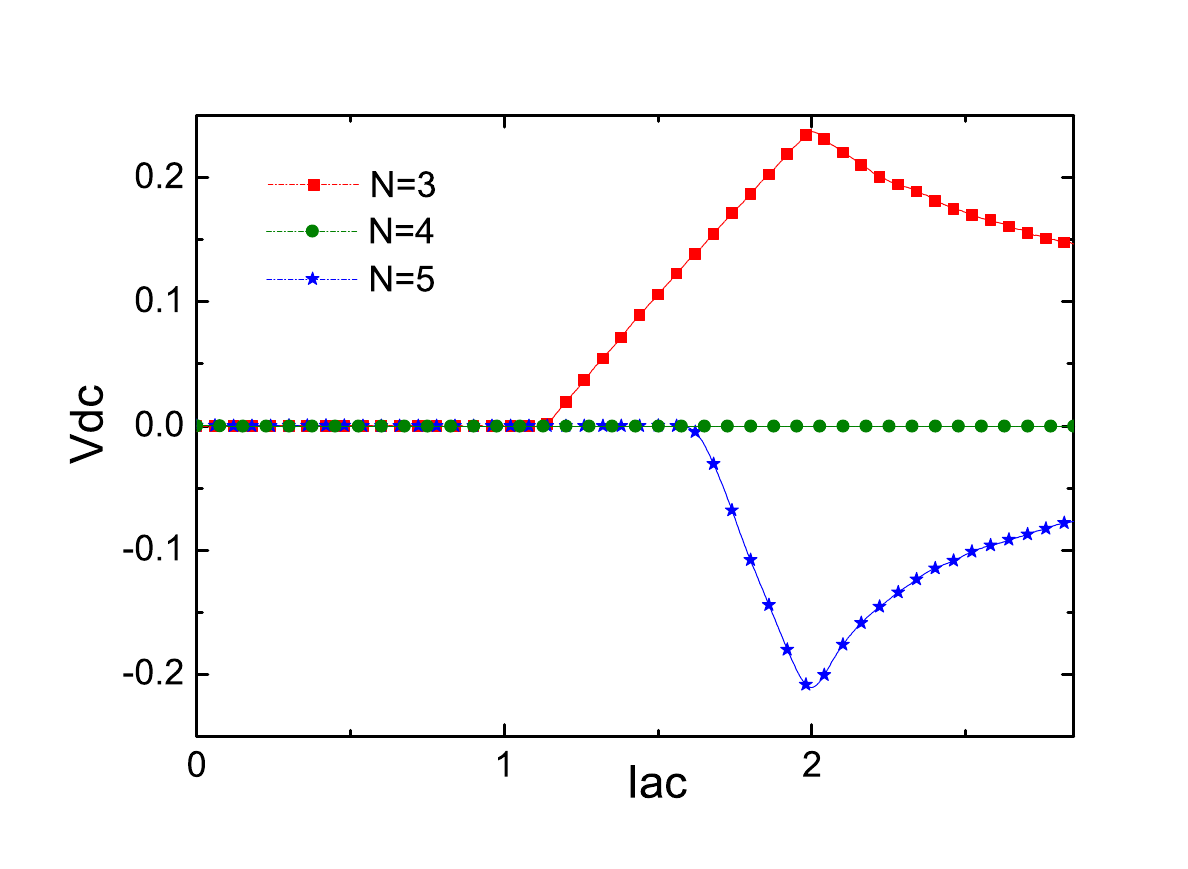}
\end{center}
\caption{Rectified voltage response obtained using RCSJ model,  as a function of  ac current amplitude  for $\beta=0.01$: $N=3$ (red $\color{red} \blacksquare\color{black})$; $N=4$  (green $\color{green}\bullet\color{black})$; $N=5$ (blue $\color{blue}\bigstar\color{black})$. For all examples $\Phi/\Phi_{0} = -1/4$.}
\label{vbeta}
\end{figure}

 In short  we found that simple loop of weak links modeled with a resistive shunted model ($RSJ$) plus thermal fluctuations is in good agreement with experimental results. The following question  naturally arises: is a ring of capacitive Josephson junctions, {\it i.e.} a capacitive and  resistive shunted model ($RCSJ$)  a suitable model to  explain the observed phenomena?. In order to get an insight into this point we include the capacitive effects in our simulations. As an illustration  in Fig.\ref{vbeta} a comparison of the mean dc voltage response $V_{dc}$ vs $I_{ac}$   among the three type of contact configurations ($N=3,4,5.$) is shown.  The results presented  were obtained at zero temperature  for a magnetic flux  $\Phi/\Phi_{0} = (2n+1)/4$. A capacitive model without thermal fluctuations is shown to be a  good aproximation  to the experimental observations if $\beta \leq 1$. In Fig.~\ref{vbeta} one $\beta$-value is shown for clarity,  $\beta=0.01$, but simulations were performed incrementing $\beta$ till unity, and similar results were obtained regarding our previous main conclusions in comparison with experimental results observed in Ref.~\cite{nele}.  

\section{CONCLUSIONS}

In summary,  
{\it at finite temperature} we predicted within a model system that the amplitude of the rectified signal depends strongly on the current contacts configuration
in the JJR, in good agreement with recent diode effects measured in  Al mesoscopic triangles \cite{nele0,nele} and  in coincidence as well with a more sophisticated model including capacitive effects. 
Our results corroborate that the clue behind the  observed rectification effects  in triangle shaped SC is indeed   the superposition of the field induced persistent current with the bias current, being these phenomena robust and strong enough in front of thermal fluctuations and for more complicated modeling of the weak links.  
The key ingredients to observe the dc voltage rectification described here under thermal fluctuations are indeed field induced
persistent currents and asymmetry induced by an off-center injection of external
currents. This recipe suggests that similar rectification effects, strongly robust under finite temperature,  
should be also present in every system with a persistent current
and an asymmetric current path,  being useful  
 to design and control newer and smaller
diode devices.

\section{Acknowledgment}
This work was supported by CONICET (PIP 112-200801-01576;  00965/10), SeCyT-UNC\'ordoba and ANPCyT  (PICTO66/2008), Argentina.
The authors specially acknowledge  A.V. Silhanek for useful suggestions and experimental images shown in Fig.~\ref{fig:jjr} and A.B. Kolton for useful  discussions. 

\bibliographystyle{elsarticle-num}

\end{document}